\documentclass{ws-procs9x6}
\usepackage[english]{babel}
\usepackage{graphicx}
\usepackage{epsfig}

\begin{document}

\def\sl#1{\slash{\hspace{-0.2 truecm}#1}}
\def\beqn{\begin{eqnarray}}
\def\eeqn{\end{eqnarray}}
\def\nn{\nonumber}
%
%

\def\eqcm{\: ,}           
\renewcommand{\d}{{\rm d}}
\newcommand{\Ta}{H_T}
\newcommand{\Tb}{\tilde{H}_T}
\newcommand{\Tc}{E_T}
\newcommand{\Td}{\tilde{E}_T}
\def\be{\begin{equation}}
\def\ee{\end{equation}}
\def\bea{\begin{eqnarray}}
\def\eea{\end{eqnarray}}
\def\ket#1{\hbox{$\vert #1\rangle$}}   
\def\bra#1{\hbox{$\langle #1\vert$}}   
\def\oneh{{\textstyle {1\over 2}}}
\def\onet{{\textstyle {1\over 3}}}
\def\smoneh{{\scriptstyle {1\over 2}}}
\def\onesix{{\textstyle {1\over 6}}}
\def\oneq{{\textstyle {1\over 4}}}
\def\treh{{\textstyle {3\over 2}}}
\def\treq{{\textstyle {3\over 4}}}
\def\oneight{{\textstyle {1\over 8}}}
\def\onesq{{\textstyle {1\over \sqrt{2}}}}

\def\Re{\hbox{\rm Re\,}}
\def\Im{\hbox{\rm Im\,}}

\def\Tr{\hbox{\rm Tr\,}}
\def\Sp{\hbox{\rm Sp\,}}
\def          
\simeq{{\ \lower2pt\hbox{$-$}\mkern-13mu \raise2pt \hbox{$\sim$}\ }}
\def\dirac#1{\slash \mkern-10mu #1}

\title{\textbf{CHIRAL-ODD GENERALIZED PARTON DISTRIBUTIONS, TRANSVERSITY AND DOUBLE
TRANSVERSE-SPIN ASYMMETRY IN DRELL--YAN DILEPTON
PRODUCTION\footnote{This research is part of the EU Integrated
Infrastructure Initiative Hadronphysics Project under contract
number RII3-CT-2004-506078.}}}
\author{\normalsize M.~\textsc{Pincetti}$^\ddagger$, B.~\textsc{Pasquini}, S.~\textsc{Boffi}}
\address{Dip. di Fisica Nucleare e Teorica, Universit\`a
degli Studi di Pavia\\ and INFN, Sezione di Pavia,\\ Pavia, 27100,
Italy\\$^\ddagger$E-mail: manuel.pincetti@pv.infn.it}

\begin{abstract}
{ Within the framework of light-cone quantization we derive the
overlap representation of generalized parton distributions for
transversely polarized quarks using the Fock-state decomposition in
the transverse-spin basis. We apply this formalism to the case of
light-cone wave functions in a constituent quark model giving
numerical results for the four chiral-odd generalized parton
distributions in a region where they describe the emission and
reabsorption of a quark by the nucleon. With the transversity
distribution obtained in the forward limit of the generalized
distribution, we provide some predictions for the double
transverse-spin asymmetry in Drell-Yan dilepton production in the
kinematics of the $\mathcal{PAX}$ experiment. }
\end{abstract}

 \bodymatter

%

%
%
\section{Introduction}
Quarks and gluons are the fundamental degrees of freedom in the
study of strong interactions by means of QCD. Nevertheless, the
objects experimentally observable are hadrons and, at present, we
have not yet a complete description of how hadrons are built up in
terms of quark and gluon fields. In hard scattering processes we
parameterize this non perturbative information through a set of
parton distribution (PDs), which are hadronic matrix elements of
bilocal products of the light-front quark and gluon field operators.
At leading twist a quark-parton model of the nucleon requires three
PDs: the unpolarized distribution, $f_1(x)$, the helicity
distribution, $g_1(x)$, and the transversity distribution, $h_1(x)$.
Whereas the first two distributions are well studied physical
quantities, both from the experimental and theoretical point of
view, so far, the last one is totally unknown. As a matter of fact,
$h_1(x)$ does not contribute to inclusive deep-inelastic scattering
because of its chiral-odd nature and is only experimentally
accessible when coupled to another chiral-odd partner in the cross
section. Notwithstanding, some theoretical activity has been
developed to calculate it and to suggest some new experimental ways
to extract $h_1(x)$ from data~\cite{barone}\,. Among all the proposed
experimental investigations, the favorite one is represented by the
double transversely polarized Drell--Yan dilepton
production~\cite{ralston}\,. In fact, this allows the study of the
double transverse-spin asymmetry of lepton-pair production, giving
direct access to the transversity distribution.

In recent years a novel class of parton distributions has been
introduced, the generalized parton distributions
(GPDs)~\cite{muller}\,. The GPDs have attracted a considerable amount
of interest, since it has been pointed out that they contain a
wealth of information about the internal structure of hadrons,
interpolating between the inclusive physics of parton distributions
and the exclusive limit of form factors\cite{goeke}\,.
These new parton distributions functions are defined as non-diagonal
hadronic matrix elements of bilocal products of the light-front
quark and gluon field operators. A complete set of quark GPDs at
leading twist include four helicity conserving, usually labeled $H$,
$E$, $\tilde H$, $\tilde E$, and four helicity flipping chiral-odd
GPDs, labeled $H_T$, $E_T$, $\tilde H_T$, $\tilde
E_T$\cite{HJ98,diehl01}\,. In the forward limit, $H$, $\tilde H$ and
$H_T$ reduce to $f_1$, $g_1$ and $h_1$, respectively.

A variety of model calculations is available for helicity conserving
GPDs. Less attention has been paid up to now to the chiral-odd
sector. So far, there is only one proposal to access the chiral-odd
GPDs in diffractive double meson production~\cite{ivanov,pire}\,. In
this paper we study the chiral-odd GPDs by means of the overlap
representation of light-cone wave functions (LCWFs) that was
originally proposed in Refs.~\cite{diehl2}\,. In a
fully covariant approach the connection between the overlap
representation of GPDs and the non-diagonal one-body density matrix
in momentum space has further been explored in Refs.~\cite{BPT03}\,
for the helicity conserving sector, making use of the correct
transformation of the wave functions from the (canonical)
instant-form to the (light-cone) front-form description. In this way
the lowest-order Fock-space components of LCWFs with three valence
quarks are directly linked to wave functions derived in constituent
quark models (CQMs).
Recently this approach has been extended to include the next-order
Fock-state component, by developing a convolution formalism for the
unpolarized GPDs which incorporate the sea quark
distribution~\cite{BP06}\,.

In this paper, following the approach discussed in
Refs.~\cite{PPB05,PPB062}\,, the four chiral-odd GPDs are derived in
Sec.~2 by means of a Fock-state decomposition in the transverse spin
basis and with LCWFs for three valence quarks. In Sec.~3 we present
the numerical results obtained for the GPDs, while we show our
estimations for the transversity distributions in Sec.~4 and some
predictions for the double transverse-spin asymmetry in Drell--Yan
dilepton production in Sec.~5.

\section{The Overlap Representation}

The chiral-odd GPDs are off-diagonal in the parton helicity basis.
In the reference frame where  the momenta $\vec p$ and $\vec p\,'$
of the initial and final nucleon lie in the $x-z$ plane, they become
diagonal if one changes basis from eigenstates of helicity to
eigenstates of transversity, i.e. states with spin projection
$\uparrow$ ($\downarrow$) directed along (opposite to) the
transverse direction $\hat x$. In such a basis they are obtained by
means of the following relations
\begin{eqnarray}
H^{q}_T &=& \frac{1}{\sqrt{1 - \xi^2}}T^q_{\uparrow\uparrow} -
\frac{2M\xi} {\epsilon\sqrt{t_0 - t}(1 -
\xi^2)}T^q_{\uparrow\downarrow},
\label{eq:accat}\\
E_T^{q} &=& \frac{2M\xi}{\epsilon\sqrt{t_0 - t}}\frac{1}{1 -
\xi^2}T_{\uparrow\downarrow}^q + \frac{2M}{\epsilon\sqrt{t_0 - t}(1
- \xi^2)}\tilde{T}^q_{\uparrow\uparrow}
\nonumber\\
&&- \frac{4M^2}{(t_0 - t)\sqrt{1 - \xi^2}(1 - \xi^2)}
\bigg(\tilde{T}^q_{\downarrow\uparrow} -
T^q_{\uparrow\uparrow}\bigg),
\label{eq:et}\\
\tilde{H}_T^{q} &=& \frac{2M^2}{(t_0 - t)\sqrt{1 -\xi^2}}
(\tilde{T}^q_{\downarrow\uparrow} - T^q_{\uparrow\uparrow}),
\label{eq:tildeaccat}\\
\tilde{E}^{q}_T &=& \frac{2M}{\epsilon\sqrt{t_0 - t}(1 -\xi^2)}
\bigg(T^q_{\uparrow\downarrow} +
\xi\tilde{T}^q_{\uparrow\uparrow}\bigg)\nonumber\\ && -
\frac{4M^2\xi}{(t_0 - t)\sqrt{1 - \xi^2}(1 - \xi^2)}
\bigg(\tilde{T}^q_{\downarrow\uparrow} -
T^q_{\uparrow\uparrow}\bigg), \label{eq:tldeet}
\end{eqnarray}
where $t=\Delta^2$ is the transferred momentum square, $- t_0 = 4
m^2 \xi^2/(1-\xi^2)$ is its minimum value for a given value of the
skewness parameter $\xi=-\Delta^+/2P^+$ defined in terms of the
momentum transfer $\Delta^\mu=p'^\mu-p^\mu$ and the average nucleon
momentum $P^\mu=\oneh(p+p')^\mu$, and $\epsilon =
\mathrm{sign}(D^1)$, where $D^1$ is the $x$-component of $D^\alpha =
P^+ \Delta^\alpha - \Delta^+ P^\alpha$.

The transverse matrix elements are defined as follows
\begin{eqnarray}
\label{eq:T} T^{q}_{\lambda'_t\lambda_t} &=& \langle p',
\lambda'_t|\int\frac{dz^-}{2\pi}e^{i \bar xP^+z^-}
\bar{\psi}(-{z}/2)\gamma^+\gamma^1\gamma_5 \psi({z}/2)|p,
\lambda_t\rangle,
\\
\label{eq:Ttilde} \tilde{T}^{q}_{\lambda'_t\lambda_t} &=&\langle p',
\lambda'_t|\int\frac{dz^-}{2\pi}e^{i\bar xP^+z^-}
\frac{i}{2}\bar{\psi}(-{z}/2)\sigma^{+1}\psi({z}/2) |p,
\lambda_t\rangle.
\end{eqnarray}
In the region $\xi \leq \bar x\leq 1$ of plus-momentum fractions,
where the generalized quark distributions describe the emission of a
quark with plus-momentum $(\bar x+\xi)P^+$ and its reabsorption with
plus-momentum $(\bar x-\xi)P^+$, they can be written in the overlap
representation of LCWFs as
\begin{eqnarray}
\label{eq:T_ov} T^{q }_{\lambda'_t\lambda_t} &=&
\sum_{N,\beta=\beta'} \bigg(\sqrt{1 - \xi}\bigg)^{2 -
N}\bigg(\sqrt{1 + \xi}\bigg)^{2 - N} \sum_{j=1}^N
\mathrm{sign}(\mu_{j}^{t})
\delta_{s_jq}\nonumber\\
&&\times\int[d\bar{x}]_N[d^2\vec{k}_\perp]_N\delta(\bar{x} -
\bar{x}_j)\Psi^{*}_{\lambda'_t,N,\beta'}(\hat{r}')
\Psi_{\lambda_t,N,\beta}(\tilde{r}),
\\
& &\nonumber\\
\label{eq:Ttilde_ov} \tilde{T}^{q }_{\lambda'_t\lambda_t} &=&
\sum_{\beta,\beta',N} \bigg(\sqrt{1 - \xi}\bigg)^{2 -
N}\bigg(\sqrt{1 + \xi}\bigg)^{2 - N}
\sum_{j=1}^N\delta_{{\mu_j^t}'-\mu_j^t}\delta_{{\mu_i^t}'\mu_i^t}\mathrm{sign}(\mu_{j}^{t})
\delta_{s_jq}\nonumber\\
&&\times\int[d\bar{x}]_N[d^2\vec{k}_\perp]_N\delta(\bar{x} -
\bar{x}_j)\Psi^{*}_{\lambda'_t,N,\beta'}(\hat{r}')
\Psi_{\lambda_t,N,\beta}(\tilde{r}),
\end{eqnarray}
where $\Psi_{\lambda_t ,N,\beta}$ is the momentum LCWF of the
$N$-parton Fock state, $s_j$ labels the quantum numbers of the
$j$-th active parton, with transverse initial (final) spin
polarization $\mu^{t}_{j}$ (${\mu_j^t}'$),  and $\mu^{t}_{i} \,
({\mu_i^t}')$ are the transverse spin of the spectator initial
(final) quarks. The set of kinematical variables $\tilde{r},
\hat{r}'$ are defined according to Refs.\cite{diehl2,PPB05}\,.

\section{Results}
As an application of the general formalism we consider the
valence-quark contribution ($N=3$) to the chiral-odd GPDs with the
LCWF derived starting from an instant-form of the proton obtained in
the relativistic quark model of Ref.\cite{Schlumpf94a}\,. The
structure of the nucleon wave function in this model is SU(6)
symmetric for the spin-isospin
components. 
\begin{figure}[ht]
\begin{center}
\epsfig{file=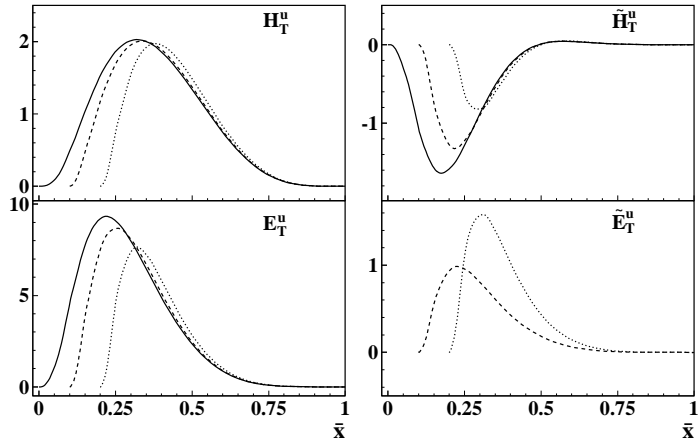,  width=22 pc}
\end{center}
\vspace{-0.4cm} \caption{\footnotesize The chiral odd generalized
parton distributions calculated in the CQM of Ref.\cite{Schlumpf94a}
for the flavour $u$
at $t=-0.2$ (GeV)$^2$ and for different values of $\xi$:
$\xi =0$ (solid curves), $\xi=0.1$ (dashed curves), $\xi=0.2$
(dotted curves).} \label{fig:fig1}
\end{figure}
As an example we show in Fig.~\ref{fig:fig1} the four calculated
chiral-odd GPDs, $H^u_T$, $E^u_T$, $\tilde H^u_T$, $\tilde E^u_T$ at
$t=-0.2$ (GeV)$^2$ and for different values of $\xi$. In all cases
the GPDs vanish at $\bar x=\xi$ since in our approach they include
the contribution of valence quarks only and we cannot populate the
so-called ERBL region with $\vert\bar x\vert\le \xi$ where
quark-antiquark pairs and gluons are important. Therefore, at low
$\bar x$ this gives a strong $\xi$ dependence of the peak position
of the distribution, but for large $\bar x$ the $\xi$ dependence
turns out to be rather weak.

Concerning the $t$ dependence, it affects the low-$\bar x$ region
and is more pronounced in the cases of $E^q_T$ and $\tilde H^q_T$.
For large $\bar x$ values the decay of all the distributions towards
zero at the boundary $\bar x=1$ is almost independent of $t$ (see
Ref.\cite{PPB05})\,.

\section{The Forward Limit}
In the forward limit $\Delta^\mu\rightarrow 0$ ($\bar x\to x$, with
$x$ being the usual Bjorken variable), only the quark GPD $H_T^q$
can be measured and, in fact, becomes the quark transversity
distribution $h_1^q(x)$. Although the quark GPDs $E_T^q$ and $\tilde
H_T^q$ do not contribute to the scattering amplitude, they remain
finite in the forward limit, whereas $\tilde E_T^q$ vanishes
identically being an odd function of $\xi$ as already noticed in
Ref.\cite{diehl01}\,.

The transversity distribution $h^q_1$ is the counterpart in the
transverse-polarization space of the helicity parton distribution
$g^q_1$ which measures the helicity asymmetry. As it was stressed by
Jaffe and Ji\cite{ralston}\,, in nonrelativistic situations where
rotational and boost operations commute, one has $g^q_1=h^q_1$.
Therefore the difference between $h^q_1$ and $g^q_1$ is a measure of
the relativistic nature of the quarks inside the nucleon. In
light-cone CQMs these relativistic effects are encoded in the Melosh
rotations.
\begin{figure}[h]
\begin{center}
\psfig{file=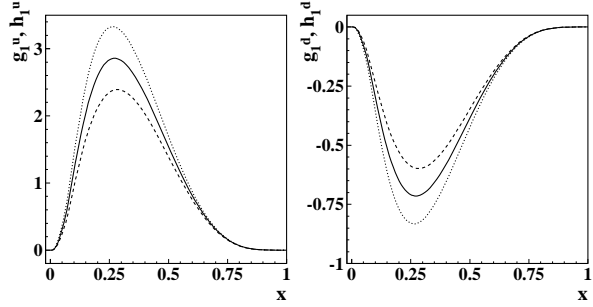,  width=19 pc}
\end{center}
\caption{\footnotesize Helicity and transversity distributions for
the $u$ (left panel) and $d$ (right panel) quark. The solid lines
correspond to $h^q_1$, the dashed lines show $g^q_1$, and the dotted
lines are the nonrelativistic results when Melosh rotations reduce
to the identity ($h^q_1=g^q_1$).} \label{fig:fig2}
\end{figure}

In Fig.~\ref{fig:fig2} the helicity and transversity distributions,
$g_1$ and $h_1$, obtained as a forward limit of the corresponding
GPDs calculated with the CQM of Ref.\cite{Schlumpf94a}\,, are
compared and plotted together with the nonrelativistic result when
Melosh rotations reduce to identity. The large difference between
$g_1$ and $h_1$ shows how big is the effect of relativity.

Recalling the expression for the unpolarized parton distribution
$f^q_1$ obtained in Ref.\cite{BPT03}\,, it is easy to see that the
following relations hold
\begin{eqnarray}
\label{eq:soffer} 2h^u_1(x)=g^u_1(x)+\frac{2}{3}f^u_1(x), & \quad&
2h^d_1(x)=g^d_1(x)-\frac{1}{3}f^d_1(x),
\end{eqnarray}
which are compatible with the Soffer inequality\cite{Soffer}\,. In
the nonrelativistic limit one obtains  $h^u_1=g^u_1=2/3 f^u_1$ and
$h^d_1=g^d_1=-1/3 f^d_1.$

\section{The Double Transverse-Spin Asymmetry}
In order to directly access transversity via Drell-Yan lepton pair
production one has to measure the double transverse-spin asymmetry
$A_{TT}$ in collisions between two transversely polarized
hadrons:\bea\label{DSA}A_{TT} = \frac{\d\sigma(\uparrow\uparrow) -
\d\sigma(\uparrow\downarrow)}{\d\sigma(\uparrow\uparrow) +
\d\sigma(\uparrow\downarrow)},\eea with the arrows denoting the
transverse directions along which the two colliding hadrons are
polarized. The most favorable situation for a sizeable effect is the
process $p^\uparrow\bar{p}^\uparrow \rightarrow l^+l^-X$ mediated by
a virtual photon\cite{Efremov}\,. At LO, i.e. considering only the
quark-antiquark annihilation graph, the double transverse-spin
asymmetry for this process is given by\bea A_{TT}^{p\bar{p}}=
a_{TT}\frac{\sum_qe^2_q[h_1^q(x_1,Q^2)h_1^q(x_2,Q^2)+h_1^{\bar{q}}(x_1,Q^2)h_1^{\bar{q}}(x_2,Q^2)]}{\sum_qe^2_q[f_1^q(x_1,Q^2)f_1^q(x_2,Q^2)+f_1^{\bar{q}}(x_1,Q^2)f_1^{\bar{q}}(x_2,Q^2)]}
,\eea where $e_q$ is the quark charge, $Q^2$ the invariant mass
square of the lepton pair (dimuon), and $\textrm{a}_{TT}$ the spin
asymmetry of the QED elementary process $q\bar{q} \rightarrow
l^+l^-$.
\begin{figure}[h]
\begin{center}
\psfig{file=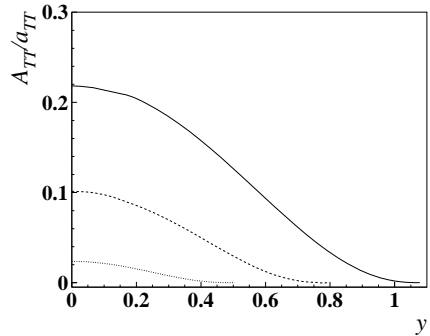,  width=14pc}
\end{center}
\caption{\footnotesize The $A_{TT}^{p\bar{p}}/a_{TT}$ at a center of
mass energy square of 45 GeV$^2$ and different values of $Q^2$: 5
GeV$^2$ (solid line), 9 GeV$^2$ (dashed line), 16 GeV$^2$ (dotted
line).} \label{fig:fig3}
\end{figure}
Here, as in Ref.\cite{PPB062}\,, we provide quantitative estimates
of $A_{TT}^{p\bar{p}}/a_{TT}$ for the kinematics of the proposed
$\mathcal{PAX}$ experiment at GSI. In Fig.~\ref{fig:fig3}\, our
predictions for $A_{TT}^{p\bar{p}}/a_{TT}$ are plotted in terms of
the rapidity $ y=1/2\ln(x_1/x_2)$, where the transversity
distribution has been suitably evolved solving numerically the DGLAP
equations\cite{A-P} (see Fig.~\ref{fig:fig4}\,)\, and where the
unpolarized distribution functions $f_1(x,Q^2)$ are taken from the
GRV parameterization\cite{GRV}\,.
\begin{figure}[h]
\begin{center}
\psfig{file=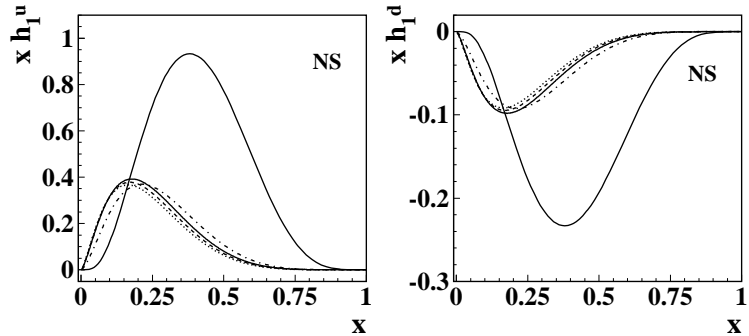,  width=24 pc}
\end{center}
\caption{ \footnotesize Evolution of the transversity distribution
for the \emph{u} (left panel) and \emph{d} (right panel) quark.
Starting from the hadronic scale of the model $Q^2_0$ = 0.079
GeV$^2$ (upper curve), we plot the LO curves at different scales:
$Q^2$ = 5 GeV$^2$, solid curves;  $Q^2$ = 9 GeV$^2$, dashed curves;
$Q^2$ = 16 GeV$^2$, dotted curves, and a NLO curve at $Q^2$ = 5
GeV$^2$ dashed-dotted curve.} \label{fig:fig4}
\end{figure}

One may notice that our predictions for $A_{TT}^{p\bar{p}}/a_{TT}$
are much lower than the results obtained in others phenomenological
analysis~\cite{Efremov}\,. However, our study suggests the
possibility of a measurable asymmetry, about 20\%, at moderate $Q^2$
values, around 5$\div$10 GeV$^2$, giving direct access to the
transversity distribution.



\end{document}